\documentclass[structabstract]{aa}
\usepackage[dvips]{graphicx}
\usepackage{txfonts}
\usepackage{color}
\usepackage{natbib}
\bibpunct{(}{)}{;}{a}{}{,} % to follow the A&A style
\usepackage{psfrag}
\usepackage{version}
\begin{document}

\title{H$_2$ formation via the UV photo-processing of a-C:H nano-particles}
\authorrunning{A.P. Jones et al.}

\author{A.P. Jones, E. Habart}
           
    \institute{Universit\'e Paris Sud and CNRS, Institut d'Astrophysique Spatiale, UMR8617, Orsay F-91405, France\\
    \email{Anthony.Jones@ias.u-psud.fr} }

    \date{Received 7 May 2015 / Accepted 26 July 2015}

   \abstract
% context
{The photolysis of hydrogenated amorphous carbon, a-C(:H), dust by UV photon-irradiation in the laboratory leads to the release of H$_2$ as well as other molecules and radicals.  This same process is also likely to be important in the interstellar medium.}
%aims
{To investigate molecule formation arising from the photo-dissociatively-driven, regenerative processing of a-C(:H) dust.}
% methods
{We explore the mechanism of a-C(:H) grain photolysis leading to the formation of H$_2$ and other molecules/radicals.}
% results
{The rate constant for the photon-driven formation of H$_2$ from a-C(:H) grains is estimated to be $2 \times 10^{-17}$\,cm$^3$\,s$^{-1}$. In intense radiation fields photon-driven grain decomposition will lead to fragmentation into daughter species  rather than H$_2$ formation.}
% conclusions
{The cyclic re-structuring of arophatic a-C(:H) nano-particles appears to be a viable route to formation of H$_2$ for low to moderate radiation field intensities ($1 \lesssim G_0 \lesssim10^2$), even when the dust is warm ($T \sim 50-100$\,K).}

\keywords{Interstellar Medium: dust, emission, extinction -- Interstellar Medium: molecules -- Interstellar Medium: general}

\maketitle

%------------------------------------------------------------------
\section{Introduction}
%------------------------------------------------------------------

The evolution of hydrocarbon solids was considered in detail in our preceding works \citep{1990MNRAS.247..305J,2009ASPC..414..473J,2012A&A...540A...1J,2012A&A...540A...2J,2012A&A...542A..98J,2013A&A...558A..62J,2014P&SS..100...26J,Faraday_Disc_paper_2014}. Here we extend that work and explore the formation of molecules arising from the vacuum ultraviolet (VUV, $E \sim 7-10$\,eV) and extreme ultraviolet  (EUV, $E \sim 10-13.6$\,eV) photon-driven processing and decompositional evolution of hydrogenated amorphous carbon particles, a-C(:H), in the interstellar medium (ISM).  The designation a-C(:H) is used to indicate the entire range of hydrogen-poor (a-C) to hydrogen-rich (a-C:H) carbonaceous solids.  The UV photolysis or ion irradiation a-C:H materials leads to their aromatisation, H atom loss and the {\it in situ} formation of molecular hydrogen \citep[{\it e.g.},][]{1984JAP....55..764S,1989JAP....66.3248A,1996MCP...46...198M,2011A&A...529A.146G,2014A&A...569A.119A}. The most recent of these experiments by \cite{2014A&A...569A.119A} show that the VUV ($E \sim 6.8-10.5$\,eV, $\lambda \sim 120-180$\,nm) photolysis of hydrogenated amorphous carbons is indeed a viable mechanism for the formation of H$_2$, and also CH$_4$, within an astrophysical framework. The photo-processing of a-C(:H) particles by VUV/EUV photons (hereafter referred to as UV photons)\footnote{The physical depth into a-C(:H) materials at which the optical depth is unity for VUV and EUV photons ($E = 7-13.6$eV) varies little with wavelength and material composition \citep[{\it e.g.},][]{2012A&A...540A...2J} and we can therefore consider the collective effects of VUV and EUV photons.} from the ambient ISRF leads to re-structuring/decomposition, H$_2$ formation and the liberation of H$_2$ and hydrocarbon daughter products \citep[{\it e.g.},][]{1984JAP....55..764S,2014A&A...569A.119A,2015MNRAS.447.1242D}. This process could also liberate (the precursors to) species such as CCH, c-C$_3$H$_2$, C$_3$H$^+$, C$_4$H, which have been observed in ISM regions with moderate UV radiation fields \citep[{\it e.g.},][]{2005A&A...435..885P,2012A&A...548A..68P,2015ApJ...800L..33G}.

In the classical picture of H$_2$ formation in the ISM an H atom physisorbed on a grain surface\footnote{The physisorption  interaction energy is $\sim 0.01-0.1$\,eV.} reacts with a second from the gas phase (Eley-Rideal mechanism) or with another on the grain surface (Langmuir-Hinshelwood mechanism) before reacting to form an H$_2$ molecule which is then ejected into the gas phase. In their recent study \cite{2014A&A...569A.100B} showed that the Langmuir-Hinshelwood mechanism on small grains with fluctuating temperatures can still be an efficient route to H$_2$ formation in unshielded photon dominated regions (PDRs). The detailed analysis of \cite{2014A&A...569A.100B} shows that surface reactions on small, stochastically-heated particles are therefore no barrier to H$_2$ formation in PDRs and can provide means to form H$_2$ where the classical mechanisms do not appear to work.
Here we describe a new mechanism for H$_2$ formation via the photolysis of stochastically-heated a-C nano-particles, which is complementary to that proposed by \cite{2014A&A...569A.100B}. In this new scenario the H atoms involved in H$_2$ molecule formation are chemically-bonded\footnote{The aliphatic C$-$H bond energy is $\sim 4.2-4.6$\,eV, while that for olefinic and aromatic C$-$H bonds is $\sim 4.8-4.9$\,eV.} within the grains, as in the experiments of \cite{2014A&A...569A.119A}. The mechanism appears to be viable route to molecule formation in PDRs with moderate-strength radiation fields where the grains are warm. 

In this paper we propose a molecule/radical formation pathway via the UV photon-driven C$-$H and C$-$C bond dissociation within, and the associated decomposition of, a-C(:H) nano-particles, which is consistent with carbonaceous dust evolution in the ISM.  The paper is organised as follows: 
Section~\ref{sect_mechanism} proposes a new mechanism for molecular hydrogen and daughter molecule formation, 
Section~\ref{sect_H2_formation} quantifies H$_2$ formation in a-C(:H) solids   
and 
Section~\ref{sect_conclusions} presents the conclusions of this work.

%------------------------------------------------------------------
\section{A mechanism for molecule formation} 
\label{sect_mechanism}
%------------------------------------------------------------------ 

% FIGURE *********************************************************
\begin{figure}
 \resizebox{\hsize}{!}{\includegraphics[angle=270]{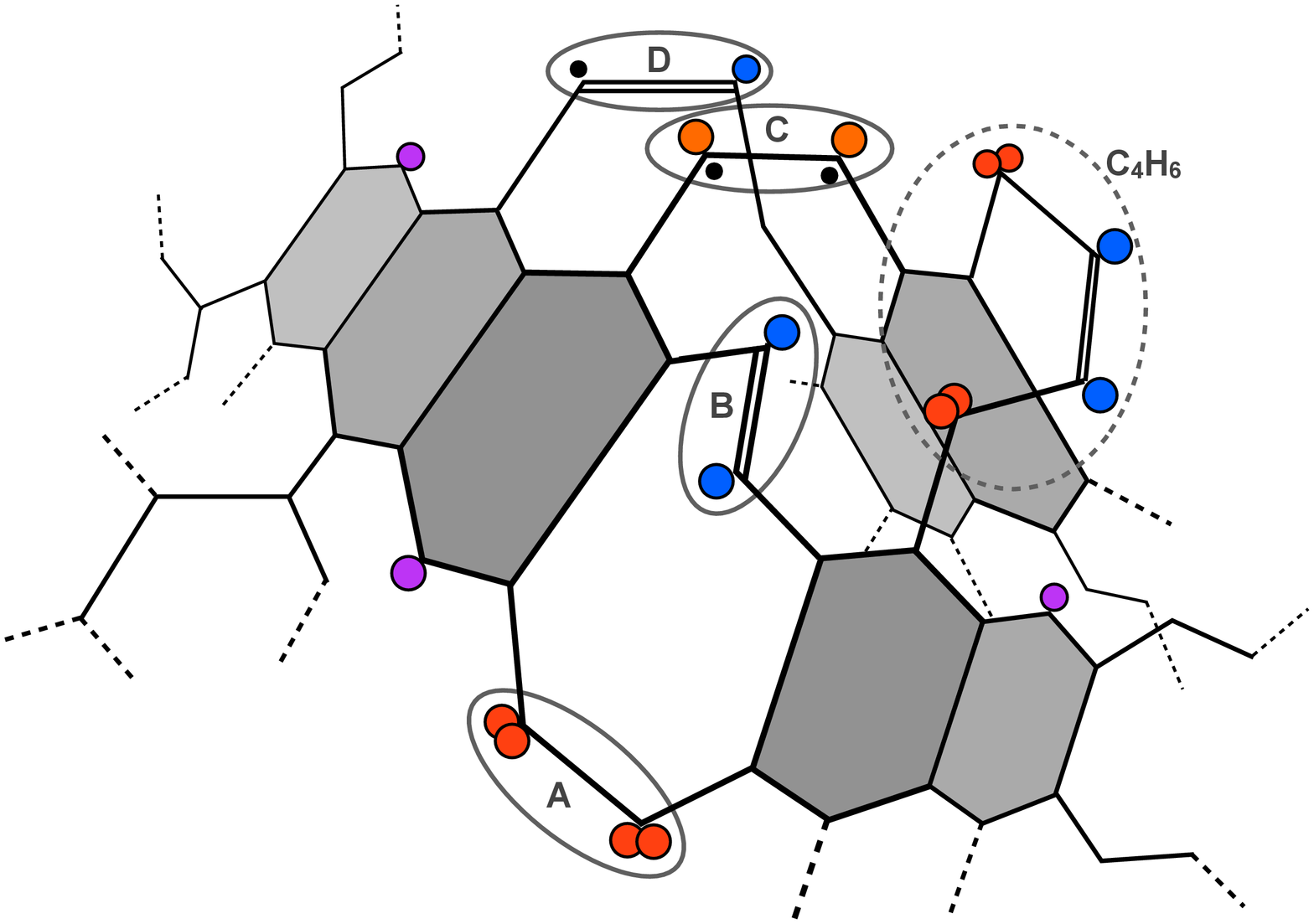}}  
 \caption{Schematic view of part of the 3D structure of an a-C(:H) grain containing $\sim 60$ C atoms and $\sim 40$ H atoms. Some of the aliphatic, olefinic and aromatic H atoms inherent within the structure are indicated by the red, blue and purple filled circles, respectively. Radical C atom or dangling bond sites sites are shown by the black dots. See text for details.}
 \label{fig_aCH_schematic}
\end{figure}
% *********************************************************

Experimental evidence shows that methyl groups, CH$_3$, within a-C:H materials probably play an important role in H$_2$ and CH$_4$ formation during UV photolysis \citep[{\it e.g.},][]{2014A&A...569A.119A}. In the diffuse ISM and PDRs it will be the small, stochastically-heated grains that dominate H$_2$ formation because of the large surface area that they provide \citep{2014A&A...569A.100B}. In the \cite{2013A&A...558A..62J} dust model the smallest grains ($a < 10$\,nm) are narrow band gap ($E_{\rm g} = 0.1$\,eV), H-poor, a-C nano-particles with no methyl groups apparent in their structure \citep{2012A&A...540A...1J,2012A&A...540A...2J,2012A&A...542A..98J}. 
Thus, in the tenuous ISM it is unlikely that methyl groups would contribute to H$_2$ formation. 

The key structural component in molecule formation from and within a-C nano-particles \cite[$a \lesssim 5$\,nm][]{2012A&A...542A..98J} would appear to be the presence of the aliphatic and olefinic bridging groups, which link the aromatic domains in these arophatic structures  \citep[{\it e.g.},][]{2012A&A...542A..98J,2012ApJ...761...35M}. Fig. \ref{fig_aCH_schematic} shows a schematic view of part of an a-C(:H) nano-particle where the aliphatic $-$CH$_2-$CH$_2-$ and olefinic $-$CH$=$CH$-$ bridges between aromatic clusters are indicated by the ellipsoidal areas A  and B, respectively. A longer  $-$C$_4$H$_6-$ mixed aliphatic-olefinic bridge is indicated by the dashed ellipse on the right of Fig. \ref{fig_aCH_schematic}. The de-hydrogenation and dissociation of such structures could be at the heart of molecule formation within a-C(:H) grains. 

H$_2$ molecule formation in a-C(:H) is somewhat analogous to that arising from the aromatic-aliphatic transformation on the periphery of PAH molecules proposed by \cite{2008ApJ...679..531R} and the catalytic formation of H$_2$ by coronene as studied experimentally by \cite{2012ApJ...745L...2M}. However, it is here hypothesised that H$_2$ formation occurs at the aliphatic/olefinic bridging structures that link the aromatic clusters in a-C(:H) in nano-particles ({\it e.g.}, the A and B sites in Fig. \ref{fig_aCH_schematic}). As \cite{2004PhilTransRSocLondA..362.2477F}, \cite{2007JChPh.126o4705H} and \cite{2007JAP...102g4311H}  point out, sp$^2$ chain and cage-like clusters may be rather common structures in a-C:H materials. Our own work indicates aromatic sp$^2$ clusters containing $3-8$ rings in a-C nano-particles with radii $\simeq 0.2-0.4$\,nm \citep{2012A&A...542A..98J,2012ApJ...761...35M}.
As Fig. \ref{fig_aCH_schematic} shows the aromatic domains and the aliphatic/olefinic bridges in these arophatic-type structures are not co-planar but rather cage-like. Hence, under the effects of CH bond UV photo-dissociation the aliphatic/olefinic (A/B) bridging structures are unable to transform into aromatic structures, or increase the aromatic domain size, without compromising the cage-like structure, {\it i.e.}, without breaking CC bonds. An aliphatic bridge (A) can, however, transform into an olefinic bridge (B), which is likely to be sterically-strained because of the distortions introduced by the random network structure. The olefinic bridge could then flip between olefinic (B) and radical aliphatic states (C) with dangling bonds, {\it i.e.}, undergo B$\Leftrightarrow$C isomerisational flipping. 
Such a mechanism is also consistent with the formation of daughter hydrocarbon radicals and molecules. 

For a simple, two-carbon atom aliphatic  bridge, A, the underlying H$_2$ formation process depends on an  equilibrium between UV photolysis and H atom addition \citep[{e.g.},][see also Section \ref{sect_H2_formation}]{2001A&A...367..347M,2001A&A...367..355M,2014A&A...569A.119A}. The mechanism for H$_2$ formation proposed here relies on the aliphatic bridges being dehydrogenated, {\it i.e.}, 
\[
-{\rm CH}_2-{\rm CH}_2- \ \rightarrow\  -{\rm CH}={\rm CH}- + \ {\rm H}_2  
\]
\vspace{-0.6cm}
\[
\hspace{0.9cm} {\rm A} \hspace{2.4cm} {\rm B}
\]
\vspace{-0.2cm}
and H atom addition being aided by the potential for sterically-hindered, de-hydrogenated aliphatic bridges to isomerise,  {\it i.e.} 
\[
-{\rm CH}={\rm CH}-  \ \leftrightarrow  \  -{\rm C}^\bullet {\rm H}-{\rm C}^\bullet{\rm H}-
\]
\vspace{-0.6cm}
\[
\hspace{0.9cm} {\rm B} \hspace{2.4cm} {\rm C}
\]
\vspace{-0.2cm}
where A, B, C, {\it etc.} also refer to Fig. \ref{fig_aCH_schematic} and the $^\bullet$ symbol is used to indicate a radical species with dangling bonds on a carbon atom that can chemisorb  H atoms. As detailed above this isomerisation scheme involves an aliphatic-to-olefinic transition that cannot proceed to aromatisation in arophatic structures because of steric limitations \citep[{\it e.g.},][]{2012A&A...542A..98J,2012ApJ...761...35M}. 

As implied above, and adopted in the following analysis, a UV photon is assumed to dissociate a single C$-$H bond releasing an H atom that can them react with a bonded H atom to form H$_2$. In the above, and for simplicity in what follows, we show adjacent H atoms forming an H$_2$ molecule. However, the reacting H atoms need not necessarily be adjacent because the C$-$H bond dissociation-released H atom can migrate through the structure and react with a more distant H atom.\footnote{Reaction with an H atom in an aliphatic C$-$H bond will be favoured because of the lower bond energy, compared to olefinic and aromatic C$-$H bonds (see footnote 2).} 

The full sequence for H$_2$ formation and aliphatic bridging structure re-cycling can be summarised by the following schematic of the UV-catalysed, H$_2$ formation cycle: 
\[
-{\rm CH}_2-{\rm CH}_2- \ \ \  [{\rm aliphatic\ bridge}] \ \rightarrow
\]
\vspace{-0.6cm}
\[
\hspace{0.9cm} {\rm A} 
\]
\[
({h\nu^\ast}:{\rm H}_2) \ \rightarrow  \ -{\rm C}^\bullet {\rm H}-{\rm C}^\bullet{\rm H}-  \leftrightarrow  -{\rm CH}={\rm CH}-
\]
\vspace{-0.6cm}
\[
\hspace{3.05cm} {\rm C} \hspace{2.25cm} {\rm B}
\]
\[
({\rm H}:\  ) \ \ \ \ \ \ \ \ \rightarrow  \ -{\rm CH_2}-{\rm C}^\bullet{\rm H}-  
\]
\vspace{-0.6cm}
\[
\hspace{2.8cm} {\rm A/C} 
\]
\[
({\rm H}:\ ) \ \ \ \ \ \ \ \ \rightarrow  \ -{\rm CH_2}-{\rm CH_2}- \ \ \  [{\rm re-generated\ aliphatic\ bridge}]
\]
\vspace{-0.6cm}
\[
\hspace{3.0cm} {\rm A} 
\]
\vspace{-0.2cm}
where we adopt the descriptor (reactant : product): with ${h\nu^\ast}$ and H being dehydrogenating UV photons and adsorbing H atoms, respectively, and H$_2$ the UV-photolysis product. The label A/C is used to indicate a mixed A and C bridge. For this H$_2$ formation cycle to operate efficiently it relies on the re-generating H atom addition time-scale being, at least, of same order of magnitude as that for the UV-photolysis step, which appears to be the case (see section \ref{sect_dust_model}). However, as the radiation field intensity increases the cycle will be increasingly interrupted by the competing dehydrogenation processes  
\[
-{\rm CH_2}-{\rm C}^\bullet{\rm H}-  \ \rightarrow \ \ ({h\nu^\ast}:{\rm H}_2) \ \rightarrow \   -{\rm CH}={\rm C}^\bullet-
\]
\vspace{-0.6cm}
\[
\hspace{0.7cm} {\rm A/C} \hspace{4.45cm} {\rm D}
\]
\[
-{\rm CH}={\rm CH}-  \ \ \ \, \rightarrow \ \  ({h\nu^\ast}:{\rm H}_2) \ \rightarrow \ -{\rm C}\equiv{\rm C}- \ ,
\]
\vspace{-0.6cm}
\[
\hspace{0.85cm} {\rm B} 
\]
\vspace{-0.2cm}
{\it i.e.}, at some point the UV-photolysed dehydrogenation becomes more important than H atom addition. In the case of extremely intense radiation fields, the UV-photolysis rate will exceed the H atom addition rate and H$_2$ formation will no longer be possible. In this case, the above processes can lead to photo-fragmentation of the structure \cite[{\it e.g.},][]{2010A&A...510A..36M,2010A&A...510A..37M,2012A&A...545A.124B,2013A&A...558A..62J}
and the formation of daughter molecule/radical species, such as C$^\bullet_2$H and C$_2$, as the particles undergo UV-photon-induced catastrophic destruction. 

If the aliphatic/olefinic bridges are longer, as shown by the dotted ellipse in Fig. \ref{fig_aCH_schematic}, {\it i.e.}, 
\\[0.2cm]
---CH$_2$---CH$=$CH---CH$_2$--- , \\[0.2cm]  
---CH$_2$---CH$_2$---CH$_2$--- \ \ or \\[0.2cm] 
---CH$_2$---CH$_2$---CH$_2$---CH$_2$--- , \\[0.2cm]
then a whole series of fragments, with the generic formula C$_n$H$_{(m \leq 2n)}$, where $n = 2,3,4$\ldots and $m=0\rightarrow 2n$,  are possible \cite[{\it e.g.}, see][]{2012A&A...540A...2J,2015MNRAS.447.1242D}. It appears that such an a-C(:H) decomposition mechanism could perhaps better explain the origin of the observed C$_2$H, C$_3$H, l-C$_3$H$^+$, c-C$_3$H$_2$, l-C$_3$H$_2$ and C$_4$H daughter products observed in the \object{Horsehead Nebula} PDR by \cite{2005A&A...435..885P,2012A&A...548A..68P} and \cite{2015ApJ...800L..33G}. These species would then originate from the UV photo-fragmentation of the aliphatic and olefinic structures that bind the aromatic clusters in a-C(:H) nano-particles, rather than from the photo-dissociation of free-flying precursor PAHs.
In the most recent of these works \cite{2015ApJ...800L..33G} show that in the \object{Horsehead Nebula} PDR the IR band emission, H$_2$ 2.12$\,\mu$m ro-vibrational line and the l-C$_3$H$^+$ emission peaks are coincident and on the leading edge of the PDR ($A_{\rm V} < 1$) just in front of the C$_2$H and c-C$_3$H$_2$ emission peaks at $A_{\rm V} \sim 1$. The latter, neutral hydrocarbon species are consistent with the break up of the a-C(:H) aliphatic/olefinic bridges in the PDR and the release of aromatic clusters into the gas that are then excited on the strongly-illuminated face of the PDR. 

Rather convincing evidence for an intimate mix of aliphatics with aromatics in carbonaceous nano-particles in the ISM (see the Fig.~\ref{fig_aCH_schematic} schematic view), as predicted by \cite{2012A&A...542A..98J} and \cite{2013A&A...558A..62J}, comes from the observed $3-4\,\mu$m region emission band spectra in many sources, {\it e.g.}, \object{NGC\,7023} \citep{2015arXiv150204941P}, the \object{Orion Bar} \citep{1997ApJ...474..735S,2001A&A...372..981V}, \object{M17\,SW}, \object{NGC\,2023} \citep{2001A&A...372..981V}, \object{M\,82} \citep{2012A&A...541A..10Y} and the Seyfert\,1 galaxy \object{NGC\,1097} \citep{2012ApJ...751L..18K}. Even in these energetic regions the 3.3\,$\mu$m aromatic CH band always appears to be accompanied by a  CH$_n$ aliphatic plateau in the $3.4-3.6\,\mu$m region. These observations are hard to explain within the context of PAH molecules in the strict chemical sense, {\it i.e.}, polycyclic aromatics with only sp$^2$ C atoms with peripheral H atoms.  The ``interstellar PAH'' model therefore requires that the IR emission band band-carriers are more complex species than  PAHs \citep[{\it e.g.},][]{2012ApJ...760L..35L} and are probably more akin to the a-C(:H) nano-particle arophatics \citep{2012ApJ...761...35M} that we propose are at the heart of H$_2$ molecule formation in PDRs (see Fig.~\ref{fig_aCH_schematic}).

%------------------------------------------------------------------
\section{The molecular hydrogen formation rate}
\label{sect_H2_formation}
%------------------------------------------------------------------

The formation of H$_2$, associated with the aliphatic to aromatic transformation of a-C(:H) dust, should also occur in the ISM where UV photons progressively de-hydrogenate and aromatise a-C(:H) materials. This coupled dehydrogenation/aromatisation process will clearly be accompanied by a de-volatilisation due to the associated loss of small molecules \cite[{\it e.g.},][]{1984JAP....55..764S,2012A&A...540A...2J,2014A&A...569A.119A,2015MNRAS.447.1242D}. Thus, the chemistry in PDRs would appear to be an ideal laboratory to search for tracers of this H$_2$ and small molecule formation mechanism, which will depend upon the radiation field intensity rather than, simply, the gas density. 

The H$_2$ formation rate in the ISM, $dn[{\rm H_2}]/dt$, is classically described by the balance between its UV photo-destruction in the gas phase and its formation by H atom recombination on the surfaces of dust particles, {\it i.e.}, 
\begin{equation}
\frac{dn[{\rm H_2}]}{dt} = -\ \beta_0 \, n_{\rm H_2} \ + \ R_{\rm H_2} \, n_{\rm p} \, n_{\rm H}, 
\label{eq_H2_equilibrium}
\end{equation}
where $\beta_0$ is the H$_2$ photo-dissociation rate coefficient ($5 \times 10^{-11}$\,s$^{-1}$) in an unshielded interstellar radiation field (ISRF), $R_{\rm H_2}$ is the H$_2$ formation rate coefficient for formation via H atom sticking and recombination on dust, $n_{\rm p}$ is the total proton density, and $n_{\rm H}$ and $n_{\rm H_2}$ are the atomic and molecular hydrogen number densities, respectively. The observationally-determined H$_2$ formation rate coefficient in the diffuse ISM, $R_{\rm H_2} \simeq 3-4 \times 10^{-17}$\,cm$^3$ s$^{-1}$ \citep[{e.g.},][]{2002A&A...391..675G}, appears to be rather invariant, indicating a degree of uniformity in the H$_2$ formation process.

\cite{2014A&A...569A.119A} recently showed that the VUV photo-processing ($\approx 8.6$\,eV per photon) of a-C:H leads to bond photo-dissociation, decomposition and the formation of H$_2$ {\em within} the a-C(:H) bulk material. The H$_2$ formation rate, $R_{\rm H_2}$, is then a first order reaction with rate constant $k_f [{\rm CH}]$ and Eq.~(\ref{eq_H2_equilibrium}) becomes 
\begin{equation}
\frac{dn[{\rm H_2}]}{dt} = -\ \beta_0 \, n_{\rm H_2} \ + \ k_f [{\rm CH}] \, n_{\rm p}  
\label{eq_H2_form_rate}
\end{equation}
and as per \cite{2014A&A...569A.119A}, in their Section 4.1, both terms on the right hand side of Eq.~(\ref{eq_H2_form_rate}), {\it i.e.}, the H$_2$ destruction and formation rates, are directly proportional to the UV photon flux.\footnote{As noted above we assume that a single UV photon results in the formation of an H$_2$ molecule.} As given by Eq.~(39) in \cite{2012A&A...540A...2J},  $k_f [{\rm CH}]$ can be expressed as 
\begin{equation}
k_f [{\rm CH}] = \Lambda_{\rm UV,pd}(a) \ N_{\rm CH}(a,d_1) \ X_i(a) \ \ \ {\rm [\,s^{-1}]}
\label{eq_H2_formation_rate}
\end{equation}
where $\Lambda_{\rm UV,pd}(a)$ is the UV photo-darkening\footnote{An increase in the {\em dark}, graphite-like or aromatic content in response to UV photon irradiation. Hence, the term photo-darkening is often used for the aromatisation of a-C(:H) materials.} rate \citep[Eq.~31][]{2012A&A...540A...2J}, $X_i(a)$ is the grain relative abundance, $N_{\rm CH}(a,d_1)$ is the number of CH bonds in a grain of radius $a$ photolysable to a depth $d_1$ ($\simeq 20$\,nm)\footnote{This is the path length corresponding to an optical depth of unity for UV photons in an homogeneous, bulk a-C(:H) material. For $E > 10$\,eV the optical depth is only weakly-dependent on the a-C(:H) material band gap and the photon energy (see paper II). It could also perhaps be regarded as the depth into the grains at which UV photolysis is in equilibrium incident H atom re-hydrogenation.} and therefore capable of forming H$_2$. Eq.~(\ref{eq_H2_formation_rate}) implicitly assumes that every absorbed UV photon leads to the photo-dissociation of a single CH bond and the formation of an H$_2$ molecule. This is almost certainly not the case because there will be competition from other channels, such as photon absorption leading to grain heating or photo-electron emission. Recent experimental measurements indicate low photo-dissociation efficiencies, {\it i.e.}, 0.014 \citep{Belen_Mate_FaradayDisc168} and 0.025 \citep{2014A&A...569A.119A} for the CH bonds in bulk a-C:H materials. Here we assume a size-dependent photo-dissociation efficiency $\epsilon = 2/a$\,[nm] for $a \geq 2$\,nm and $\epsilon = 1$ for $a < 1$\,nm \citep[see][for further details]{2012A&A...540A...2J,2012A&A...545C...2J,Faraday_Disc_paper_2014}.\footnote{With this definition of $\epsilon$ the experimentally-derived values of 0.014 and 0.025 \citep{Belen_Mate_FaradayDisc168,2014A&A...569A.119A} are equivalent to $a \simeq 140$ and 80\,nm, respectively, which can be considered as bulk materials in terms of their optical properties \citep{2012A&A...542A..98J}.} Following \cite{2012A&A...540A...2J,2012A&A...545C...2J} the UV photo-darkening rate, and hence the H$_2$ formation rate, is given by 
\begin{equation}
\Lambda_{\rm UV,pd}(a) = F_{\rm UV} \ \sigma_{\rm CH-diss.} \ Q_{\rm abs}(a,E) \, \epsilon\ \ \ {\rm [\,s^{-1}]},  
\label{eq_UV_dehyd_rate}
\end{equation}
where $F_{\rm UV} \simeq 3 \times 10^7$\,photons cm$^{-2}$ s$^{-1}$ is the CH bond-dissociating UV photon flux in the local ISM \citep{2002ApJ...570..697H} for $G_0 =1$, $\sigma_{\rm CH-diss.} \simeq 10^{-19}$\,cm$^2$ is the CH bond photo-dissociation cross-section for $E_{\rm h \nu} \gtrsim 10$\,eV
\citep{1972JChPh..57..286W,1994CPL...227..243G,2001A&A...367..355M,2001A&A...367..347M,2014A&A...569A.119A} and $Q_{\rm abs}(a,E)$ is the particle absorption efficiency factor in the cross-section, {\it i.e.}, $\sigma(a,E) = \pi a^2 Q_{\rm abs}(a,E)$.  The a-C(:H) particle photo-processing time-scale, as a function of grain radius and the radiation field, can be expressed as 
\begin{equation}
\tau_{\rm UV,pd}(a,G_0) = \frac{1}{\Lambda_{\rm UV,pd}(a) \, G_0}. 
\label{eq_tau_UVpd}
\end{equation}

From Eq.~(\ref{eq_H2_form_rate}) the molecular hydrogen fraction, at equilibrium ($dn[{\rm H_2}]/dt = 0 $),  is
\[
\frac{n_{\rm H_2}}{n_{\rm H}} =  \frac{k_f [{\rm CH}]}{\beta_0} %\ G_0} 
\]
\begin{equation}
\ \ \ \ \ =  \left[ \frac{F_{\rm UV} \, \sigma_{\rm CH-diss.}}{\beta_0} \right] \Bigg\{ Q_{\rm abs}(a,E) \, \epsilon(a) \,N_{\rm CH}(a,d_1) \, X_i(a) \Bigg\}.  
\label{eq_H2_equilib_frac}
\end{equation}
\cite{2014A&A...569A.119A} derived a similar expression for this ratio (see in their Section 4.2) but did not include the size-dependent optical property terms for the dust. 

Within the framework of the proposed model, the UV photon flux is therefore key in determining the H$_2$ formation and destruction rates, which scale with the local ISRF intensity and hardness. However, even if H$_2$ formation via a-C:H UV photolysis is viable, it will be hampered by the limited stock of H atoms within the dust, which is determined by the carbon depletion into dust, {\it i.e.}, ${\rm [C/H]} \approx 10^{-4}$. As shown by \cite{2012A&A...540A...2J}, H$_2$ formation would be sustainable for only $\approx 10^4$\,yr before all CH bonds were photo-dissociated and the grains completely de-hydrogenated during aromatisation. An a-C re-hydrogenation mechanism,  such as that proposed by \cite{2008ApJ...682L.101M,2010ApJ...718..867M}, is therefore needed to replenish the stock of CH bonds. Thus, the complete dehydrogenation/aromatisation of a-C(:H) grains must be hindered and suitably stable and numerous H atom chemisorption sites need to be preserved. These sites could be dangling bonds resulting from steric limitations on the $sp^3$ to $sp^2$ transformation (see Section \ref{sect_mechanism}) or due to the low-level doping of a-C(:H) with nitrogen atoms \citep[{\it e.g.},][see Section \ref{sect_nitrogen}]{1991JAP....70.4958A,2013A&A...555A..39J}. In this case, H$_2$ formation would be sustainable as long as the radiation field is intense enough to photo-dissociate CH bonds but not intense enough to break the CC bonds in the aliphatic/olefinic bridging structure and hence to photo-fragmentation the a-C(:H) grains.  Thus, this H$_2$ formation mechanism will not work at high extinction because there are too few UV photons, nor will it work in intense radiation fields because of the rapid photo-fragmentation of the grains, {\it i.e.}, their destruction via the formation and ejection of daughter molecules/radicals \citep[{\it e.g.},][see also Section~\ref{sect_daughter_mols}]{1984JAP....55..764S,2005A&A...435..885P,2012A&A...540A...2J,2013A&A...558A..62J,2014A&A...569A.119A}. These ideas appear to be in general agreement with the conclusions of \cite{2004A&A...414..531H,2011A&A...527A.122H}, who find enhanced H$_2$ formation and higher than expected column densities of rotationally excited H$_2$ in moderately-excited PDRs.

%------------------------------------------------------------------
\subsection{A role for nitrogen heteroatoms?}
\label{sect_nitrogen}
%------------------------------------------------------------------

Apparently, hetero-atom groups, {\it i.e.}, chemical groups containing atoms other than C or H, are of low abundance or are absent from carbonaceous materials in the ISM \citep{2002ApJS..138...75P,2005A&A...432..895D}. Nevertheless, N atoms were clearly detected in the comet Halley CHON carbonaceous phase \cite[in CHON and CHN combinations,][]{1987A&A...187..779C}  indicating N hetero-atoms within solar system carbonaceous dust. 

Interestingly, a-C:H materials can be rather inefficiently doped with nitrogen atoms to form a-C:H:N materials \citep{1994DRM.....3.1034S} and this could have some interesting consequences for the observable properties of a-C(:H) dust in the ISM and circumstellar regions \citep{2013A&A...555A..39J,2014P&SS..100...26J}. \cite{1991JAP....70.4958A} find that low N-doped a-C:H materials (N concentration $<1$ atomic \%), with an optical gaps $\simeq 1$\,eV, are rich with dangling bonds\footnote{A dangling bond is an unsatisfied valence on an atom in a chemically-bonded network, which is an active chemisorption site.} at a concentration of $5 \times 10^{20}$\,cm$^{-3}$. N heteroatom-induced dangling bonds could provide H atom trapping sites for the catalytic formation of H$_2$ in N-doped arophatics. 

The inclusion of N atoms into amorphous carbons to form amorphous carbon nitrides, a-CN$_x$, has led to the classification of a number of N-bonding configurations \cite[{\it e.g.},][]{2004PhilTransRSocLondA..362.2477F}.
In their experimental study \cite{2005A&A...432..895D} found that the nitrogen atoms in their laboratory interstellar a-C:H analogues tend to be found in structure-bridging groups rather than in aromatic clusters. However, the similar frequencies for CÐC and CÐN modes can make the interpretation of the skeletal modes difficult \cite[{\it e.g.},][]{2004PhilTransRSocLondA..362.2477F}.

% FIGURE *********************************************************
\begin{figure*}
\begin{center}
 \resizebox{15.0cm}{!}{\includegraphics{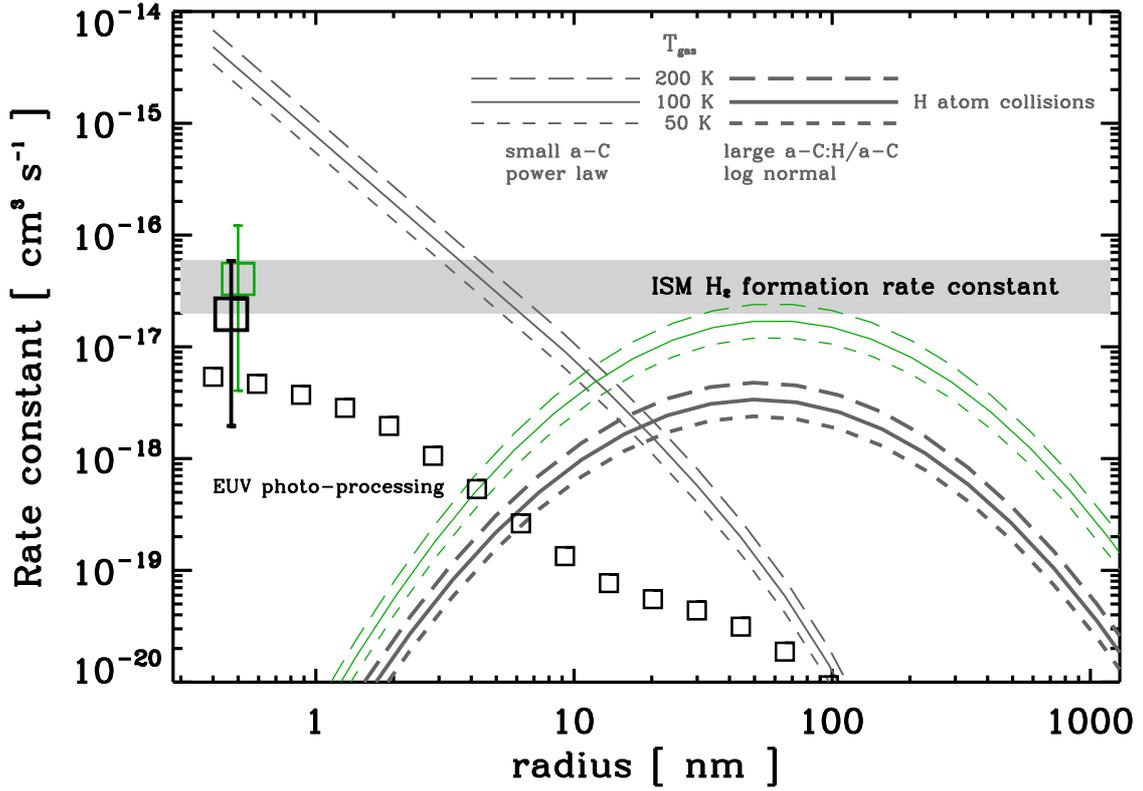}}
 \caption{ The canonical interstellar H$_2$ formation rate constant \cite[grey horizontal band,][]{2002A&A...391..675G} compared to that predicted  for H$_2$ formation via CH bond UV photo-dissociation from a-C:H/a-C grains, $k_f$[CH]/$n_{\rm H} = 2.0 \times 10^{-17}$\,cm$^3$\,s$^{-1}$  shown at $a=0.5$\,nm (squares with error bars).  The green green square also includes H$_2$ formation in the a-C mantles on the large a-Sil$_{\rm Fe}$ grains into the calculation, yielding $k_f$[CH]/$n_{\rm H} = 4.0 \times 10^{-17}$\,cm$^3$\,s$^{-1}$. The small squares show the relative contributions to the total H$_2$ formation rate as a function of particle radius. The H atom sticking rate constant, for $s_{\rm H} = 1$, is shown for the small a-C (thin grey), large a-C:H/a-C (thick grey) and a-Sil/a-C (green) grain distributions and for gas temperatures of 50, 100 and 200\,K (short-dashed, solid and long-dashed, respectively).}
 \label{fig_H2_formation_rate_const}
\end{center}
\end{figure*}
% *********************************************************

%------------------------------------------------------------------
\subsection{The dust model and H$_2$ formation}
\label{sect_dust_model}
%------------------------------------------------------------------

In this work we adopt the dust compositions and size distributions from the \cite{2013A&A...558A..62J}/\cite{2014A&A...565L...9K} dust model, which is specifically-tuned to the evolution of the optical and thermal properties of a-C(:H) grains in the ISM \citep[{\it i.e.}, using the optEC$_{\rm (s)}$ and optEC$_{\rm (s)}$(a) data from][]{2012A&A...540A...2J,2012A&A...542A..98J,2012A&A...545C...2J,2012A&A...545C...3J}. The \cite{2013A&A...558A..62J} model consists of a power-law distribution of small a-C grains, with a bulk material band gap $E_{\rm g} = 0.1$\,eV, and a log-normal distribution of a-C:H/a-C (core/mantle) grains, with $E_{\rm g} = 2.5$ \,eV and 0.1\,eV, respectively. These hydrocarbon populations are complemented by a log-normal distribution of large amorphous forsterite- and pyroxene-type silicate grains, containing iron and iron sulphide nano-particle inclusions, and with a 5\,nm thick coating of a-C \citep[a-Sil$_{\rm Fe}$/a-C,][]{2013A&A...558A..62J,2014A&A...565L...9K}. The model, strongly-constrained by laboratory-data, is able to explain the observed variations in the IR-FUV extinction, the 217\,nm UV bump, the IR absorption and emission bands  and the IR-mm dust thermal emission.

The H atom collision and sticking rate coefficient for a grain of radius $a$ and relative abundance $X_i(a)$ is given by 
\begin{equation}
R_{\rm H} = \pi a^2 \ X_i(a) \ \left( \frac{8 \, k T_k}{\pi \, m_{\rm H}} \right)^{\frac{1}{2}} \ s_{\rm H}, 
\label{eq_H_stick_rate}
\end{equation}
where $T_k$ is the gas kinetic temperature and $s_{\rm H}$ is the H atom sticking efficiency. Fig.~\ref{fig_H2_formation_rate_const} shows this rate coefficient, $R_{\rm H}/n_{\rm H}$, assuming $s_{\rm H} = 1$, for gas temperatures of 50, 100 and 200\,K.  

The $n_{\rm H}$-normalised, UV photon-induced, H$_2$ formation rate coefficients, $k_f [{\rm CH}]/n_{\rm H}$, integrated over the small and large a-C(:H) grain size distributions are indicated in Fig.~\ref{fig_H2_formation_rate_const} by the large squares with error bars. These data give an good fit to the observationally-derived H$_2$ formation rate coefficient in the diffuse ISM, {\it i.e.}, $R_{\rm H_2} \simeq 3-4 \times 10^{-17}$\,cm$^3$ s$^{-1}$ \citep{2002A&A...391..675G}, which is indicated by the horizontal grey band.  

The size-dependent a-C(:H) particle photo-darkening or CH bond photo-dissociation time-scale in the diffuse ISM was recently estimated by \cite{Faraday_Disc_paper_2014}. They found that the a-C(:H) photo-processing time-scale  can be analytically expressed as a function of the grain radius, $a$, in nm;  
\begin{equation}
\tau_{\rm UV,pd}(a) = \frac{10^4}{G_0} \bigg\{ 2.7  + \frac{6.5 }{(a\,[{\rm nm}])^{1.4}} + 0.04 \, (a\,[{\rm nm}])^{1.3}  \bigg\} \ \ {\rm [yr]},     
\label{eq_pd_timescale}
\end{equation} 
which implies a minimum processing time-scale, of the order of $(4 \times 10^4/G_0)$\,yr, for  particles with $a \sim 3-10$\,nm and processing time-scales $\geqslant (10^5/G_0)$\,yr for the smallest and largest grains in the assumed a-C(:H) particle size distribution \citep{Faraday_Disc_paper_2014}. Here we assume the particle size-dependent, photo-darkening efficiency factor, $\epsilon(a)$ given above (see Section \ref{sect_H2_formation}). For a-C(:H) grains such a size-dependent CH bond photo-dissociation efficiency appears to be a physically realistic because with increasing particle size other channels will compete with CH photo-dissociation,  {\it e.g.}, UV photon absorption leading to thermal excitation and/or fluorescence.  However, the proposed H$_2$ formation mechanism is only viable if the H atom sticking rate is higher than the CH photo-dissociation rate and Fig.~\ref{fig_H2_formation_rate_const} shows that this is indeed the case. We find an H$_2$ formation rate constant of $2.0 \times 10^{-17}$\,cm$^3$\,s$^{-1}$ for a-C:H/a-C grains and $4.0 \times 10^{-17}$\,cm$^3$\,s$^{-1}$ if the a-C coated amorphous silicate grains, a-Sil/a-C, are also included. The error bar on the square data points in Fig.~\ref{fig_H2_formation_rate_const} conservatively reflects a factor of three uncertainty in the dust optical properties and a factor of ten uncertainty in $\epsilon$.

For the small a-C grains the bulk material band gap of 0.1\,eV ($\equiv X_{\rm H} \sim 0.02$) required by the \cite{2013A&A...558A..62J} model implies that these particles cannot be completely aromatised to a graphite-like material with a band gap close to 0\,eV. Small a-C grains are therefore able to preserve a residual hydrogenation fraction of a few percent that is seemingly hard to remove in the ISM without destroying and/or fragmenting the structure. They are therefore more resistant to UV photo-processing than expected \citep{2012A&A...540A...2J}, consistent with the H$_2$ formation mechanism proposed in Section \ref{sect_mechanism}. 

In addition to CH bond photo-dissociation by UV photons it is also possible that H$_2$ formation could be driven by thermal desorption during small grain, whole-particle stochastic heating events. However, as discussed by \cite[][Section 5.1]{2012A&A...540A...2J} the effects of a-C(:H) dust thermal processing may not be important. Here we do not consider thermal desorption-driven H$_2$ formation and so our predicted formation rates may only be lower limits. Given that the small a-C grain population is sensitive to photo-processing \citep[{\it e.g.},][]{2012A&A...542A..69P,2013A&A...558A..62J}, the more resistant large a-C(:H) and a-Sil$_{\rm Fe}$/a-C grains ought to provide a `background' H$_2$ formation rate. 

In their study \cite{2004A&A...414..531H} found that in moderately excited PDRs ($G_0 = 10^2-10^3$) the H$_2$ formation rate is about five times the standard value but that in higher radiation field PDRs the standard value of $3 \times 10^{-17}$\,cm$^3$ s$^{-1}$ is sufficient.  Based on their observations they also inferred a link between H$_2$ formation and the aromatic emission band carriers. In a later study \cite{2011A&A...527A.122H} found that the observed column densities of rotationally excited H$_2$ in PDRs ($G_0 = 5-10^2$) are higher than that predicted by PDR models. This discrepancy could be explained by an enhancement in the H$_2$ formation rate in PDRs \citep{2011A&A...527A.122H}. The observational evidence therefore seems to point towards PDRs as particularly efficient H$_2$-forming regions when the local radiation field is moderate, which appears to be in qualitative agreement with the model proposed here.  

The \cite{2013A&A...558A..62J} dust model predicts an enhanced abundance of small a-C grains in PDRs due to the photo-fragmentation of larger a-C(:H) grains, which is consistent with the findings of \cite{2008A&A...491..797C} and \cite{2012A&A...541A..19A}. This implies an enhanced H$_2$ formation rate in moderate PDRs where the radiation field is not intense enough destroy a-C nano-particles. In very intense radiation fields ($G_0 > 10^3$) the destruction of the emission band carriers is observed \citep[{\it e.g.},][]{1994A&A...284..956B}. Here we assume that the a-C(:H) nano-particles responsible for the IR emission bands are also the source of H$_2$ formation, and that both processes are driven by UV photon absorption.  Interestingly, the work of \cite{2012A&A...542A..69P} shows that the photo-fragmentation of very small carbonaceous grains, which are PAH-based in their model, occurs in regions where H$_2$ rotational line emission is observed. Nevertheless, extremely intense radiation fields that lead to the photo-fragmentation and destruction of a-C(:H) nano-particles \citep{2013A&A...558A..62J} will close the door on H$_2$ formation and also the IR band emission in such regions.  Hence, the absolute value of $G_0$ is a key parameter in the proposed H$_2$ formation mechanism, which is likely to be efficient over a wide range of the $G_0/n_{\rm H}$ parameter space, {\it e.g.}, $0.01 < G_0/n_{\rm H} < 1.0$ unless the local ISRF is extremely intense, {\it i.e.},  $G_0 \gg 10^3$ \citep{1994A&A...284..956B}.

%------------------------------------------------------------------
\subsection{Daughter-molecule formation}
\label{sect_daughter_mols}
%------------------------------------------------------------------

The photolysis and associated aromatisation of a-C:H materials yields decomposition products other than molecular hydrogen (Section \ref{sect_H2_formation}). However, the other released species contain at least one carbon atom and result in decomposition and a structural re-adjustment to a lower hydrogen and carbon content. \cite{1984JAP....55..764S} and \cite{2014A&A...569A.119A} showed that a-C(:H) loses its hydrogen as H$_2$ but also as small hydrocarbon molecules. Thus, a-C(:H) photolytic decomposition  leads to significant mass loss from the material during thermal- or photo-processing. It is likely that the release of daughter radical or molecular species will be more pronounced in smaller particles \cite[{\it e.g.},][]{1996MNRAS.283..343D,1997ApJ...490L.175S,2000ApJ...528..841D,2009ASPC..414..473J}.  \cite{2005A&A...435..885P} proposed that the small hydrocarbon molecules that they observed in PDR regions ({\it i.e.}, CCH, c-C$_3$H$_2$, C$_4$H)\footnote{To this list can now be added the ion C$_3$H$^+$, which has also been observed in the Horsehead Nebula \citep{2012A&A...548A..68P}.} derive from pre-exisitng PAHs, which are undergoing UV photon-induced destruction. A similar hypothesis was proposed by \cite{1996ApJ...472L.123S} and extended by \cite{2009ASPC..414..473J}, who postulated that the precursors for the `PAHs' are themselves derived from the UV photolytic de-construction of small a-C(:H) particles in the transition regions between molecular clouds and PDRs. Such a process is supported by the laboratory work of \cite{1984JAP....55..764S}, \cite{2014A&A...569A.119A} and \cite{2015MNRAS.447.1242D}, which indicate that small molecules are formed during the degradation of  bulk a-C(:H) materials. As shown by \cite{2015MNRAS.447.1242D} the primary a-C:H photo-decomposition products  are hydrocarbon species (molecules and radicals) with less than four carbon atoms, {\it e.g.}, C$_3$H$_2$, C$_3$H$_4$  and C$_n$H$_{2n-1}$ alkyl radicals. The experimental evidence therefore favours a top-down photo-processing route to small hydrocarbon formation in PDRs (see Sect. \ref{sect_mechanism}). However, \cite{2015MNRAS.447.1242D} also note the release of C$_6$H$_5$ (phenyl) and C$_6$H$_6$ (benzene) in their a-C(:H) photo-decomposition experiments, indicating that the loss of single six-fold ring systems (the smallest possible aromatic domains) also needs to be included into the aliphatic/olefinic bridge-breaking and a-C(:H) re-structuring mechanism proposed in Sect. \ref{sect_mechanism}.  

To be viable an interstellar dust model must therefore be consistent with both the PDR observations \citep{2005A&A...435..885P,2012A&A...548A..68P,2015ApJ...800L..33G} and the laboratory-observed UV photolytic-formation of small aliphatic-rich species from carbonaceous materials \citep[{\it e.g.},][]{1984JAP....55..764S,1996ApJ...472L.123S,2014A&A...569A.119A,2015MNRAS.447.1242D}. In our recent work we used the optical and physical properties of a-C(:H) materials within the framework of a new ISM core/mantle dust model, which is qualitatively-consistent with these data \citep{2009ASPC..414..473J,2012A&A...540A...1J,2012A&A...540A...2J,2012A&A...542A..98J,2013A&A...558A..62J}. In this model a-C(:H) nano-particles (with a number of C atoms $N_{\rm C} \lesssim$ a few hundred) are of mixed aliphatic/aromatic (arophatic) composition and exhibit `cage-like' cluster structures (see Fig.~\ref{fig_aCH_schematic}). Such structures have been proposed as fullerene precursors \citep{2012ApJ...761...35M} and are similar to the locally aromatic polycyclic hydrocarbons (LAPHs) proposed as IR emission band carriers by \cite{2003ApJ...594..869P}. The photon-driven fragmentation of such nano-particles in PDRs will disaggregate them into their their sub-component aliphatic/olefinic bridging structures (with $N_{\rm C} \lesssim 4$) and small aromatics \cite[with few rings, {\it i.e.}, $N_{\rm R} = 3-8$,][]{2012A&A...542A..98J}. All of these daughter species will be more rapidly photo-dissociated in PDRs than the parent nano-particle. They are therefore transient species signalling the demise of a-C(:H) nano-particles in intense radiation fields \citep{2013A&A...558A..62J} but at the same time they are the drivers of some interesting top-down interstellar chemistry. 

In PDRs it is also possible that the a-C(:H) grains are undergoing a sort of slow combustion by reaction with the abundant gas phase atomic oxygen, {\it i.e.}, 
\begin{equation}
{\rm O_{(g)}} \ + \ {\rm [a-C(:H)]C_{(s)}} {\rm } \ \ \rightarrow \ \ {\rm [a-C(:H)]} \ + \ {\rm CO_{(g)}^\uparrow}.
\label{eq_aCH_combustion}
\end{equation}
Such a process will lead to the formation of excited CO molecules in regions where the a-C(:H) dust is being eroded.

%------------------------------------------------------------------
\section{Conclusions}
\label{sect_conclusions}
%------------------------------------------------------------------

We have shown that the UV-UV photo-processing of the aliphatic/olefinic bridging structures in arophatic a-C(:H) grains could provide a viable route to H$_2$ formation in the ISM, where the ISRF is sufficiently intense, {\it i.e.}, $ 1 \leqslant G_0 \lesssim 10^2$. 

A photon-driven mechanism for H$_2$ and daughter hydrocarbon radical and molecule formation is proposed, which operates through the chemisorption of H atoms at activated sites and CC bond destruction, respectively. 
It is also possible that this H$_2$ formation mechanism could be aided by the dangling bonds favoured by low N atom doping concentrations within a-C(:H) nano-particles. 
The proposed process is not directly dependent on the grain temperature and therefore would provide a plausible route to H$_2$ formation in PDRs where the {\em large  grains}  are warm ({\it i.e.}, $T_{\rm dust} > 25$\,K) and where experiments indicate that H$_2$ formation, involving H atom physisorption on grain surfaces, 
is not efficient. We note that even in the diffuse ISM ($G_0 = 1$) the nano-particles responsible for the bulk of the H$_2$ formation, from UV photon-induced CH bond dissociation, and also for the IR emission features have temperatures of the order of $50-70$\,K. 
In more intense radiation fields these same nano-grains will have temperatures well in excess of $100$\,K. 
In PDRs where the {\em large grains} are much warmer ($T_{\rm dust} \gtrsim 50$\,K), such as the Orion region, the proposed mechanism will no longer operate because the small a-C grains that dominate H$_2$ formation are destroyed. 

In high extinction regions, the UV flux will be insufficient to drive H$_2$ formation and in extremely intense radiation fields H$_2$ formation will be severely limited because the active sites are destroyed as the grains undergo UV photon-induced fragmentation. It therefore appears that, for a given region of the ISM, there will be an optimum set of physical conditions that will favour H$_2$ formation by the proposed mechanism. 
The most favoured conditions would appear to be moderate densities and radiation fields but these will need to quantified by detailed modelling. 
Seemingly, PDR regions are the ideal sites for H$_2$ formation from the photolytic processing of a-C(:H) grains. 

%%%%%%%%%%%%%%%%%%%%%%%%%%%%%%%%%%%%%%%%%

\begin{acknowledgements} 
The authors would like to thank Emmanuel Dartois, Nathalie Ysard and Laurent Verstraete for their thoughtful and helpful remarks on the manuscript. 
%This research was, in part, made possible through the financial support of the Agence National de la Recherche (ANR) through the program
%s Cold Dust (ANR-07-BLAN-0364-01) and CIMMES (ANR-11-BS56-029-02).
\end{acknowledgements}

%%%%%%%%%%%%%%%%%%%%%%%%%%%%%%%%%%%%%%%%%

% for the bibliography, at the end
\bibliographystyle{aa} % style aa.bst
\bibliography{biblio_HAC} % your references Yourfile.bib

%\Online

%%%%%%%%%%%%%%%%%%%%%%%%%%%%%%%%%%%%%%%%%%

%% APPENDICES
%\appendix
%\newpage

%-----------------------------------------------------------------------------------------------------------------
%\newpage
%\clearpage
%\section{Title}
%\label{appendix_}

\listofobjects

\end{document}